\documentclass[twocolumn,pre,superscriptaddress,aps,showpacs]{revtex4}

\usepackage[dvips]{graphicx}
\usepackage{subfigure}
\usepackage{epsfig}

\def\be{\begin{equation}}
\def\ee{\end{equation}}
\def\bea{\begin{eqnarray}}
\def\eea{\end{eqnarray}}

\begin{document}
\title{Applicability of the hydrodynamic description of classical
  fluids}
\author{James P. \surname{Mithen} }\email{james.mithen@physics.ox.ac.uk}
\affiliation{Department of Physics, Clarendon Laboratory, University of Oxford, Parks Road, Oxford OX1 3PU, UK}
\author{ J\'er\^ome \surname{Daligault}}
\affiliation{Theoretical Division, Los Alamos National Laboratory, Los Alamos, NM 87545}
\author{Gianluca Gregori}
\affiliation{Department of Physics, Clarendon Laboratory, University of Oxford, Parks Road, Oxford OX1 3PU, UK}

\date{\today}

\begin{abstract}
We investigate using numerical simulations the domain of applicability of the hydrodynamic description of
classical fluids at and near equilibrium.
We find this to be independent of the degree of many-body correlations in the system; the range $r_c$ of the 
microscopic interactions completely determines
the maximum wavenumber $k_{max}$ at which the hydrodynamic description is applicable
by $k_{max} r_c \simeq 0.43$.  
For the important special case of the Coulomb
potential of infinite range, we show that the ordinary hydrodynamic description is never valid.
\end{abstract}


\maketitle
The equations of hydrodynamics \cite{Landau}, such as the Navier-Stokes equations, are certainly the most widely used framework for investigating the dynamics of fluids, including gases \cite{Mansour}, liquids \cite{Scopigno}, plasmas \cite{Gedalin} and nuclear matter \cite{Bouras}. 
Despite widespread use and successes, a number of outstanding questions of both fundamental 
and practical importance remain regarding the conditions under which the hydrodynamic description holds.

One ordinarily thinks of the hydrodynamic picture as applying only for wavenumbers $k$ such that $kl_f \ll 1$ with $l_f$ the mean free path and frequencies
$\omega$ such that $\omega/\omega_{c} \ll 1$ with $\omega_c$ the mean collision frequency.
These conditions, derived and already rather qualitative for a system governed by uncorrelated binary collisions (e.g. a dilute gas), become even more indeterminate when many-body correlations are present (as is the case, for example, in a liquid) because the concepts of mean free path and mean collision time cease to have a clear physical meaning.  
Thus the applicability of the hydrodynamic description certainly depends strongly on the thermodynamic conditions - e.g. the density $n$ and temperature $T$ - as well as the strength and range of the particles' mutual interactions.
For instance, one expects that the description never applies on lengthscales smaller than the range $r_{c}$ of the potential
(i.e. $k r_c \gtrsim 1$) 
-- in other words, that the domain of validity will shrink as the range increases
(such long range potentials are of particular importance in plasma physics).
In fact, in the extreme case of $r_c = \infty$, the very existence of a hydrodynamic description is a known but unsolved problem \cite{BausHansen}.
As well as leading to a deeper understanding of the emergence 
of macroscopic behaviour in interacting many-body systems, it is also of significant practical importance to know exactly
when hydrodynamics can be used to describe the behaviour of fluids, e.g. for analysing 
light and neutron scattering experiments \cite{Scopigno,BalucaniZoppi,Glenzer}.

In this Letter, we address with numerical simulations the question of the domain of applicability of the hydrodynamic description for fluids at or near equilibrium as the level of many-body
correlations in the system is varied. To this end, we consider a one-component system with a Yukawa interaction
potential $v(r) = q^2\exp(-r/r_c)/r$, where $q^2$ is the
strength and $r_c$ the range of
the potential.  Although not possessing the short range features of conventional pair
potentials used to describe normal liquids, the Yukawa potential
is certainly suitable for investigating the long lengthscale dynamics that concern
the hydrodynamic description. Additionally, it is commonly used in describing the 
screened interactions in plasmas \cite{Donkorev,SaizWunsch}.
What is more, since for $r_c = \infty$ one recovers the Coulomb
potential, we are able to use this model to answer the question of the
existence of the hydrodynamic limit referred to previously.
Most importantly however, this model is very convenient here
because it is known to be fully characterised by two
dimensionless parameters only \cite{Donkorev} -   the reduced
range $r_c^{*} = r_c/a$, where $a=\left(4\pi\/n/3\right)^{-1/3}$  is 
the average inter-particle spacing, and the
coupling strength $\Gamma = q^2/(ak_bT)$, which itself characterises completely the degree
of many-body correlations present in the system for a given range \cite{Ichimaru}.
For a wide range of $\Gamma$ and $r_c^{*}$ values, thus spanning
states ranging from dilute gases to dense liquids \cite{Daligault}, 
with short or long range microscopic interactions,
we determine the length and time scales at which
the hydrodynamic description breaks down.

To accomplish this, we have computed with Molecular Dynamics (MD) simulations the dynamical
structure factor, $S(k,\omega)$, that is the Fourier transform in
space and time of the density
autocorrelation function, for a wide range of $\Gamma$ ($0.1$,$1$,$5$,$10$,$50$,$120$,$175$) and
$r_c^{*}$ ($0.5$,$1$,$1.4$,$2$,$3.3$,$10$,$\infty$) values.  $S(k,w)$ contains complete information of the
system dynamics at and near thermal equilibrium through the Fluctuation-Dissipation
Theorem and is routinely measured in inelastic light and neutron scattering experiments 
(e.g. \cite{Scopigno,Glenzer2,Glenzer,HansenMcdonald}).  
Three main difficulties are involved with MD calculation of $S(k,\omega)$.
Firstly, for long range potentials (large $r_c^{*}$), it is essential to include 
the Ewald summation; we compute this for all our $r_c^{*}$ values
using the Particle-Particle-Particle-Mesh method \cite{Hockney}.
Secondly, obtaining accurate MD data for $S(k,\omega)$ requires averaging
the results of a large number of simulations to improve statistics. 
Thirdly, in order to investigate
the long wavelength dynamics that concern the hydrodynamic description, 
very large scale simulations (a large number of particles $N$) are needed - the minimum reduced wavevector, $(ka)_{min}$,
at which the system dynamics can be determined using MD is $\propto N^{-1/3}$.
These computational demands have made a thorough study like ours impractical before now.
In our computation of $S(k,\omega)$, we average the results of fully $25$ simulations, each 
of duration $819.2 \omega_p^{-1}$ (the plasma frequency $\omega_p =
\sqrt{3q^2}{ma^3}$ is the natural timescale for our system, where $m$ is the particle
mass), with up to $500,000$ particles.
Our complete analysis will be detailed in a forthcoming publication \cite{Mithen}.

\begin{figure}[h!]
\includegraphics{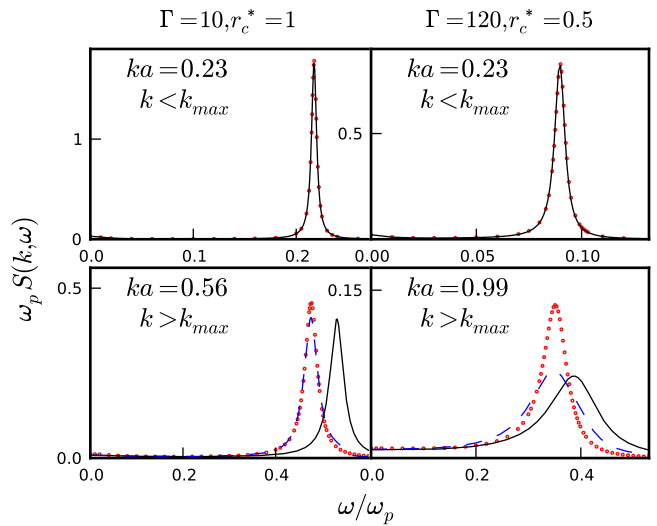}
\caption{(color online) A sample of our MD results for $S(k,\omega)$ (red dots) against $S^{H}(k,\omega)$ 
  in Eq. (\ref{hydroskw}) (black line) and when a `mean field' is added (blue dashed line - bottom panel only).}
\label{fig1}
\end{figure}

\begin{figure}[t]
\includegraphics{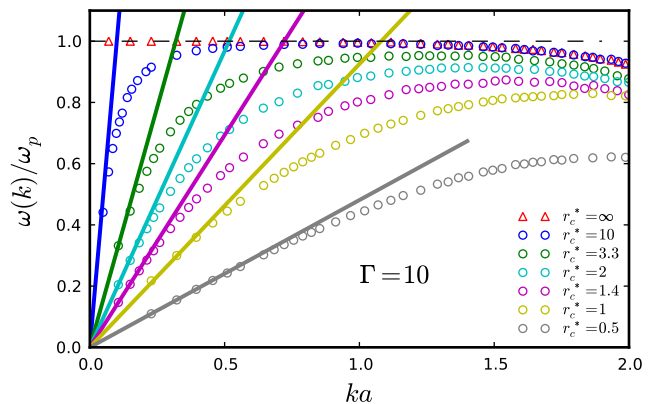}
\caption{(color online) Brillouin peak position $w(k)/\omega_p$ as obtained from MD  
(open symbols), along with the corresponding linear relations $\omega = c_s k$ (solid lines).}
\label{fig2}
\end{figure}

Firstly, we consider the case of finite range interaction potentials
($r_c^{*} < \infty$).  In this case the MD data can be compared 
to the result
 obtained from the linearised hydrodynamic (Navier-Stokes) equations 
\cite{BoonYip,HansenMcdonald}
\begin{eqnarray}
&&\frac{S^{H}(k,\omega)}{S(k)} = \frac{\gamma - 1}{\gamma}\frac{2D_Tk^2}{\omega^2 +
  (D_Tk^2)^2} \label{hydroskw}\\
&+& 1/\gamma\left[\frac{\sigma k^2}{(\omega + c_sk)^2 + (\sigma
    k^2)^2} + \frac{\sigma k^2}{(\omega - c_s k)^2 + (\sigma
    k^2)^2}\right]\nonumber \,,
\end{eqnarray}
where the static structure factor $S(k)$ in Eq. (\ref{hydroskw}) is also determined from our MD simulations.
Eq. (\ref{hydroskw}) consists of a central (Rayleigh) peak representing a diffusive
thermal mode and two Brillouin peaks at $\omega = \pm c_s k$ corresponding to propagating sound waves.
As illustrated in the top panel of Fig. \ref{fig1}, at the smallest $k$ value accessible to our MD simulations we find that the MD $S(k,\omega)$ can always
(i.e. for all $\Gamma$ and $r_c^{*}$) be very accurately fitted to Eq. (\ref{hydroskw}) , thus 
giving numerical values for the thermal
 diffusivity $D_T$, sound attenuation coefficient $\sigma$, adiabatic sound speed
$c_s$ and ratio of specific heats $\gamma$ that appear in the hydrodynamic description.
When obtained in this way, these parameters are found to be in very good
agreement with previous equation of state and transport coefficient 
calculations for the Yukawa model \cite{HamaguchiSaigoDonko}.
In particular, we find that $\gamma \approx 1$ - that is, the Rayleigh peak 
at $\omega = 0$ in Yukawa fluids is negligible.
In all cases, however, we find two Brillouin peaks, at $\omega = \pm c_s k$, representing a damped sound wave.
Fig. \ref{fig2} shows the position of the Brillouin peak obtained from our MD simulations.
We see that as the interaction potential becomes more long
ranged, it is necessary to look at increasingly long lengthscales
(small $ka$ ) for the hydrodynamic description to be applicable.
Clearly in all cases,
at some $k$ value, which we denote by $k_{max}$, the position $\omega(k)$ of the Brillouin peak as computed
by MD diverges from the linear relation.

Quantitatively, we define $k_{max}$ as the minimum $k$ value for which $\omega(k)/(c_s k) > 1.01$.
Using this criterion, for all values of the coupling $\Gamma$, we find that $k_{max}r_c \simeq 0.43$.

\begin{figure}[t]
\includegraphics[width=\columnwidth]{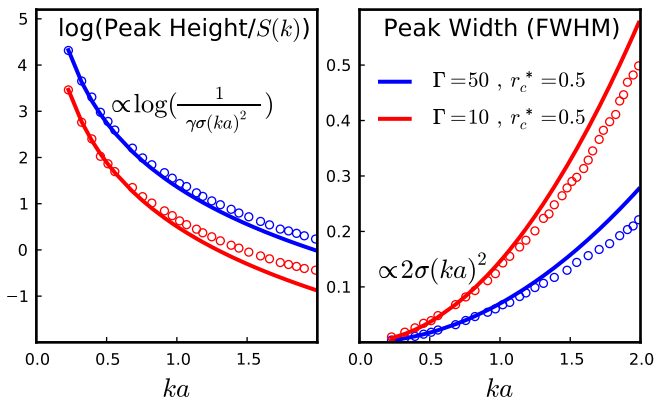}
\caption{(color online) Height and width of Brillouin peak as computed
from MD (open circles), and the predictions of Eq. (\ref{hydroskw}) (solid lines).
}
\label{fig3}
\end{figure}

The $k_{max}$ obtained from the peak position is found to also characterise well the departure of the 
height and width of the Brillouin peak from the predictions of the hydrodynamic description (Fig. \ref{fig3}).
Therefore, $k_{max}$ is the maximum wavevector at which the hydrodynamic description of Eq. (\ref{hydroskw}) is applicable.
As shown in Fig. \ref{fig3}, beyond $k_{max}$ the height of the Brillouin peak decreases more slowly, and its width
increases more slowly, than predicted by Eq. (\ref{hydroskw}).  Clearly however, the hydrodynamic
description is valid for a relatively large range of $k$ values, well beyond $k = 0$.
In real space, we find that the lengthscale $2\pi/k_{max}$ is for all
$\Gamma$ values greater than
the short-range correlation length over which the pair correlation function $g(r)$
exhibits peaks and troughs \cite{BausHansen}.
It is remarkable that $k_{max}$ does not depend on $\Gamma$; indeed, one would intuitively expect the domain
of validity of Eq. (\ref{hydroskw}) to increase as the system becomes more `collisional' (i.e. with increasing $\Gamma$).
We also note that providing $k < k_{max}$, the hydrodynamic approximation 
of Eq. (\ref{hydroskw}) for $S(k,\omega)$ is extremely accurate for all $\omega$ where $S(k,\omega)$ 
is not negligibly small; in this range, the Brillouin peaks exhaust the frequency sum-rules 
(see the top panel of Fig. \ref{fig1}).

\begin{figure}[t]
\includegraphics[width=\columnwidth]{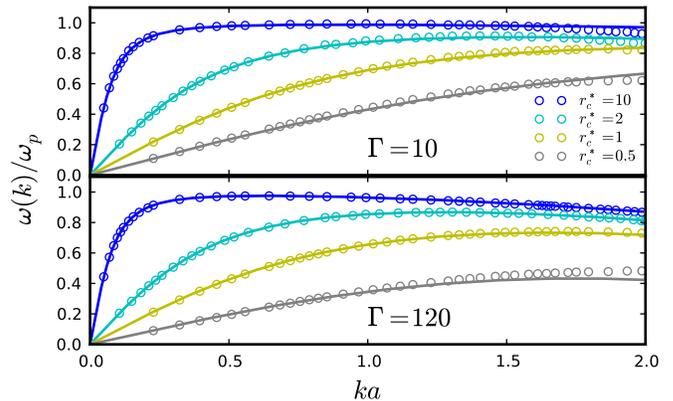}
\caption{(color online)  Brillouin peak position as obtained from MD (open circles),
and the prediction of  Eq. (\ref{salin}) (solid lines).
}
\label{meanfield}
\end{figure}

Much detailed work has been carried out to extend from macroscopic to microscopic lengthscales the domain
in which ordinary hydrodynamics applies (e.g. \cite{HansenMcdonald,BoonYip,BalucaniZoppi}).  
Interestingly, we find that simply by adding to the usual stress tensor the mean field term 
one can account very well for the position of the Brillouin 
peak.  
Microscopically, this additional term
stems from the inclusion of a self-consistent `mean field' or `Vlasov' term - usually neglected 
because one considers lengthscales longer than the range of the potential - in the appropriate kinetic equation.
By including the mean field term in the macroscopic equations, one obtains for the Yukawa model a modified expression
for the position of the Brillouin peak \cite{Salin}
\begin{equation}
\omega(k) = \left( K + \frac{\omega_p^2}{k^2 + 1/r_c^2} \right)^{\frac{1}{2}}k \,,
\label{salin}
\end{equation}
where $K = c_s^2 - \omega_p^2 r_c^2$.  We note that for systems with $\gamma = 1$, which is 
a good approximation for the $\Gamma$ and $r_c^{*}$ values considered here, the addition of the
mean field does not change the hydrodynamic description of the height or width of the Brillouin peak 
(see \cite{Salin} for details).  As shown in Fig. \ref{meanfield}, Eq. (\ref{salin}) 
 gives a remarkably good description of the Brillouin peak position, even up to $ka \approx 2$ in most cases
(although as shown in Fig. \ref{fig1} the height and width of the
peak does not always compare well with MD simulations).
Indeed, this dramatic improvement is somewhat unexpected, since dynamics at these large wavevectors 
are not usually thought to be well described by macroscopic approaches.  

For finite $r_c^{*}$, the mean field only begins to play a role when 
$kr_c > 0.43$, i.e. when the range of the potential is large compared to
the lengthscale of the density variations.
Therefore one may expect that for $r_c^{*} = \infty$,
when the interaction potential is Coulombic \cite{OCPnote},
the mean field is important at all lengthscales (in this case, our criterion $k_{max}r_c \simeq 0.43$ gives $k_{max} = 0$ !).  To be sure, 
the peculiarity of the Coulomb potential is very well known - 
in this case the longitudinal waves are not low frequency sound waves as for $r_c^{*} < \infty$ but instead high frequency plasma
waves ($\omega \approx \omega_p$), even at $k = 0$.
The resulting `plasmon' peak in $S(k,\omega)$, the position of which is illustrated in Fig. \ref{fig2} (red triangles), is certainly
not described by the low-frequency hydrodynamic
equations that lead to Eq. (\ref{hydroskw}) - one 
can indeed wonder why hydrodynamics should describe plasma oscillations at all.
The difficulty here is underlined by a kinetic theoretical derivation of the hydrodynamic equations \cite{BausHansen}: 
when proceeding with the Chapman-Enskog expansion of the appropriate kinetic equation, 
the mean field term is usually treated as a small perturbation since in the small-gradient region of interest 
to hydrodynamics the kinetic equation is always dominated by the collision term.
In this case, however, the mean field term cannot be considered as small since its straightforward 
small-gradient expansion diverges with the characteristic Coulomb divergence (see \cite{BausHansen}
and references therein).  
Based on this analysis, Baus and Hansen \cite{BausHansen} argued that only when
the collisionality dominates the mean field, which they predicted would occur at sufficiently
high coupling strength $\Gamma$, could a hydrodynamic
description be expected.
In this case the hydrodynamic description is identical to Eq. (\ref{hydroskw}) \cite{gammanote} but with $c_s k$ 
replaced with $\omega_p (1 + \frac{c_s^2k^2}{2\omega_p^2})$ \cite{Vieillefosse}.
This macroscopic description is known not to work for small $\Gamma$ \cite{BausHansen};
exactly how large $\Gamma$ has to be for it to be applicable
was left as an open question until now.  We show here that in fact the hydrodynamic description is not valid
at any $\Gamma$.

Baus and Hansen \cite{BausHansen} based the arguments outlined above on 
an exact formula for $S(k,\omega)$, derived using generalised kinetic theory  
, which at small $k$ is 
given by \cite{gammanote} 
\begin{eqnarray}
\frac{S^{B}(k, \omega)}{S(k)} &=&
  \frac{bk^2}{(\omega + \omega_p(1+\frac{k^2}{2}a))^2 + 
(\frac{k^2}{2}b)^2} \nonumber \\
&+& \frac{bk^2}{(\omega - \omega_p(1+\frac{k^2}{2}a))^2 + 
(\frac{k^2}{2}b)^2}
\,,
\label{ocpfull}
\end{eqnarray}
where   $a$ and $b$ are generalised coefficients with $k$
and $\omega$ dependence.  They were able to show that only at large $\Gamma$ would it be possible for these 
coefficients to equal their 
macroscopic counterparts (of Eq. (\ref{hydroskw})), $c_s^2/\omega_p^2$ and $2\sigma$ respectively \cite{BausHansen}.
We have estimated $a$ and $b$ by fitting $S(k,\omega)$ 
at the smallest $k$ value 
accessible to our MD simulations to Eq. (\ref{ocpfull})
 - this gives a very good fit.
However, as shown in Table \ref{ocptable}, the values obtained for 
$a$ and $b$ do not agree at all with their macroscopic counterparts, even at
our highest coupling strength of $\Gamma = 175$, which is close to the freezing point of the system \cite{BausHansen}.  
For example, the width of the plasmon peak $b$
does not even follow the same trend predicted by the hydrodynamic
scaling $2\sigma$ at our higher $\Gamma$ values.
From this we conclude that the combination of mean field and collisional effects means that
the hydrodynamic description \`{a} la Navier Stokes is not valid for a Coulomb system at any coupling strength $\Gamma$.

\begin{table}[t]
\begin{ruledtabular}
\begin{tabular}{ccccc}
$\Gamma$ & $a$ & $c_s^2/\omega_p^2$ & $b$ & $2\sigma$ \\\hline
1  & 0.895	& 0.304 & 0.192 & 1.825 \\ \hline
5  & 0.088	& -0.034 & 0.109 & 0.333 \\ \hline
10 & -0.009 & -0.080 & 0.078 & 0.235 \\ \hline
50 & -0.056 & -0.112 & 0.032 & 0.212 \\ \hline
120 & -0.062 & -0.127 & 0.021	& 0.349 \\ \hline
175 & -0.065 & -0.129 & 0.009 & 0.550 \\ \hline
\end{tabular}
\end{ruledtabular}
\caption{Comparison of the generalised coefficients obtained by
  fitting the MD $S(k,\omega)$ at our smallest $k$ value to Eq. (\ref{ocpfull}) 
  with the analogous coefficients that appear in the hydrodynamic description. 
  $c_s$ is determined from the internal energy fit given in
  \cite{DeWitt}, while $\sigma$
  is obtained from \cite{Daligault}.}
\label{ocptable}
\end{table}

In summary, for finite range potentials, $r_c^{*}< \infty$ , we find that the hydrodynamic
description is (i) always valid at sufficiently long lengthscales
where `sufficiently long' is determined by the range of the potential ($k_{max}r_c \simeq 0.43$) (ii)
extremely accurate  at these long lengthscales for all $\omega$ where $S(k,\omega)$ is not negligibly small (iii) not
enlarged in its applicability as the level of many-body correlations
in the system (i.e. $\Gamma$) is increased and (iv) is significantly extended in
applicability by including a `mean field' term in the macroscopic equations.
For a Coulomb system, $r_c^{*} = \infty$, although the macroscopic approach correctly predicts the plasmon peak at $k = 0$, for $k > 0$ the persistence of both mean field and collisional 
effects causes the ordinary hydrodynamic approach to fail.

This work was supported by the John Fell Fund at the University of Oxford and by EPSRC
grant no. EP/G007187/1.  The work of J.D. was performed for the U.S. Department of Energy
by Los Alamos National Laboratory under Contract No.
DE-AC52-06NA25396.


\begin{thebibliography}{}
\bibitem{OCPnote} We note also that for the Coulomb potential, the system is required to have
a uniform (inert) neutralising background.
\bibitem{gammanote} here we assume $\gamma = 1$ for simplicity;
this does not affect our discussion and is also a good approximation.
\end{thebibliography}
\end{document}